\documentclass[journal]{IEEEtran}
\usepackage{epsfig}

\usepackage{cite}
\ifCLASSINFOpdf
\else
\fi
\usepackage{amsmath}
\usepackage{bm}
\usepackage{array}
\usepackage{url}
\begin{document}
\title{Analysis of magnetization loss on a twisted superconducting strip in a constantly ramped magnetic field}

\author{
Yoichi~Higashi,~Huiming~Zhang,~and~Yasunori~Mawatari% <-this % stops a space
\thanks{Y. Higashi and Y. Mawatari are with National Institute of Advanced Industrial Science and Technology (AIST), Tsukuba, Ibaraki, 305-8568 Japan (e-mail: y.higashi@aist.go.jp).}% <-this % stops a space
\thanks{H. Zhang is with China Electric Power Research Institute, No.15 Xiaoying East Road, Haidian District, Beijing 100192 China}% <-this % stops a space
\thanks{Manuscript received April 8, 2018; revised August 30, 2018; accepted September 28, 2018. Date of publication October 8, 2018.}
}


\markboth{IEEE TRANSACTIONS ON APPLIED SUPERCONDUCTIVITY,~VOL.~29, NO.~1, JANUARY~2019}%
{Shell \MakeLowercase{\textit{et al.}}: Bare Demo of IEEEtran.cls for IEEE Journals}



\maketitle

\begin{abstract}
Magnetization loss on a twisted superconducting (SC) tape %tape wire
in a ramped magnetic field is theoretically investigated
through the use of a power law for the electric field--current density characteristics and a sheet current approximation.
First, the Maxwell equation in a helicoidal coordinate system is derived to model a twisted SC tape, %tape wire,
taking account of the response to the perpendicular field component in the steady state.
We show that a loosely twisted tape can be viewed as the sum 
%summation
of a portion of tilted flat tapes of infinite length by examining the perpendicular field distribution on a twisted tape.
%Next, the
The analytic formulae for both magnetization and loss power in the tilted flat tape approximation are verified based on the analytic solution of the reduced Maxwell equation
in the loosely twisted tape limit of $L_{\rm p}\rightarrow \infty$ with the twist pitch length $L_{\rm p}$.
These analytic formulae show that
%these parameters
both magnetization and loss power
decrease by a factor of $B(1+1/2n,1/2)/\pi$ (where $B$ is the beta function)
 for an arbitrary power of SC nonlinear resistivity $n$,
compared with those in a flat tape of infinite length.
%Moreover, we
%numerically
%find that
%The loss power per unit length is independent of $L_{\rm p}$ in the case of a loosely twisted tape.
Finally, the effect of the field-angle dependence of the critical current density $J_{\rm c}$ on the loss power is investigated,
and we demonstrate that it is possible to obtain an approximate estimate of the loss power value via $J_{\rm c}$ in an applied magnetic field perpendicular to the tape surface (i.e., parallel to the $c$ axis).
\end{abstract}


% \fi
\IEEEpeerreviewmaketitle

%%%%%%%%%%%%%%%%%%%%%%%%%%%%%%%%%%%%%%%%%%%%%%%%%%%%%%%%%%%%%%%%%%%%
%%%%%%%%%%%%%%%%%%%%%%%%%%%%%%%%%%%%%%%%%%%%%%%%%%%%%%%%%%%%%%%%%%%%
\section{Introduction}
%%%%%%%%%%%%%%%%%%%%%%%%%%%%%%%%%%%%%%%%%%%%%%%%%%%%%%%%%%%%%%%%%%%%
%%%%%%%%%%%%%%%%%%%%%%%%%%%%%%%%%%%%%%%%%%%%%%%%%%%%%%%%%%%%%%%%%%%%
\IEEEPARstart{S}{uperconducting} (SC) coils based on rare earth--barium--copper oxide superconductor tapes %tape wires
are being developed for use in magnetic resonance imaging (MRI) machines that operate at high magnetic fields of around 3 T \cite{yokoyama2017}.
%Inside a SC coil,
In a tape, %wire,
the screening current flows on a wide surface with a tape-shaped geometry owing to the perpendicular component of the external field induced by the transport current, 
resulting in a sizable irregular field and loss power
in the case of excitation/demagnetization of an MRI magnet.

%In order to generate a spatiotemporally uniform magnetic field inside a MRI magnet,
To reduce the thermal load of a refrigerator or remove the quench generation sources,
practical structured tapes % tape wires
with low loss are desired.
Cabling methods involving multifilamentarization \cite{amemiya2004} and/or twisting \cite{takayasu2012} are
%developed
known to be effective for reducing the magnetization loss on the tape %tape wire
%to suppress the influence of the screening current
while maintaining a high critical current density $J_{\rm c}$ in a high magnetic field.
Detailed numerical analysis of electromagnetic fields has previously been carried out for twisted multifilamentary coated superconductors in an ac magnetic field \cite{amemiya2006}.

This study addresses the macroscopic electromagnetic response in ramping magnetic fields
supposing excitation/demagnetization of an MRI magnet.
%during excitation/demagnetization of an MRI magnet.
In the case of multifilamentary tapes, %tape wires,
each of the SC filaments is electromagnetically coupled through the normal conductor embedded between the SC filaments and/or the surrounding stabilizer 
if the frequency of the ac magnetic field
exceeds a certain critical value \cite{amemiya2006}, giving rise to coupling loss.
This coupling loss can be reduced by shortening the effective wire length by twisting and electrically decoupling the current loop.
In an ac magnetic field, the time scale of the electromagnetic coupling of SC filaments
is known to be proportional to $L^2_{\rm p}$, where $L_{\rm p}$ is the twist pitch length \cite{wilson1983},
and hence a reduction in the loss power on twisted multifilamentary tapes %tape wires
upon decreasing $L_{\rm p}$ can also be expected in ramping fields.
Here, we disregard the effect of the multifilamentarization and
focus on only the effect of twisting on the macroscopic electromagnetic responses 
such as magnetization and loss power on a single SC strip.
Twisting a flat tape is beneficial to reduce magnetization loss just because it reduces the average perpendicular component of the magnetic field penetrating the tape.
The theoretical study of a twisted SC strip in a ramping field
is expected to be valuable for the development of high-field SC coil systems for MRI.

We obtain the same analytic formulae for both magnetization and loss power in two different ways;
one is the tilted flat tape approximation in which a twisted strip is regarded as
%the summation of tilted flat tape
the sum of a portion of tilted flat tapes of infinite length 
and the other is based on the Maxwell equation in the loosely twisted tape limit of $L_{\rm p}\rightarrow \infty$.
In this limit, any small correction of the higher order of $k=2\pi/L_{\rm p}$ is negligible.
Both magnetization and loss per unit length are analytically and numerically shown not to depend on $L_{\rm p}$.
Furthermore, they become smaller than those for a flat tape by a geometric factor owing to twisting alone.
A numerical analysis incorporating the field and field-angle dependence of $J_{\rm c}$ is also performed to demonstrate how to obtain
an appoximate numerical estimate of the loss power value.
%a rough numerical estimate of the loss power value.

%%%%%%%%%%%%%%%%%%%%%%%%%%%%%%%%%%%%%%%%%%%%%%%%%%%%%%%%%%%%%%%%%%%%
%%%%%%%%%%%%%%%%%%%%%%%%%%%%%%%%%%%%%%%%%%%%%%%%%%%%%%%%%%%%%%%%%%%%
\section{Model for twisted superconducting tape}
%\section{Model for twisted superconducting tape wire}
%%%%%%%%%%%%%%%%%%%%%%%%%%%%%%%%%%%%%%%%%%%%%%%%%%%%%%%%%%%%%%%%%%%%
%%%%%%%%%%%%%%%%%%%%%%%%%%%%%%%%%%%%%%%%%%%%%%%%%%%%%%%%%%%%%%%%%%%%
%%%%%%%%%%%%%%%%%%%%%%%%%%%%%%%%%
%%%%%%%%%%%%%%%%%%%%%%%%%%%%%%%%%
\begin{figure}[tb]
\centering
\includegraphics[width=80mm]{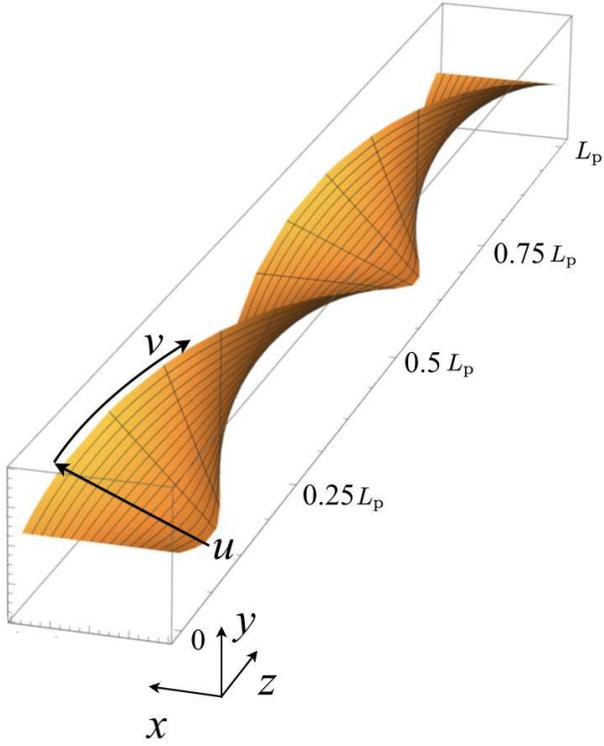}
\caption{Schematic of a twisted SC tape %tape wire
surface with a fixed twist pitch length $L_{\rm p}$.
The axes ($u,v$) are set along the tape surface.}
\label{fig1}
\end{figure}
%%%%%%%%%%%%%%%%%%%%%%%%%%%%%%%%%
%%%%%%%%%%%%%%%%%%%%%%%%%%%%%%%%%
We consider a single twisted SC tape with the tape width $w_0$ and the tape thickness $d_0$.
We then approximately regard a twisted SC tape as the twisted tape surface of an infinitesimal thickness, as shown in figure~\ref{fig1}.
The coordinates of a twisted SC tape %tape wire
can be expressed in terms of $(u,\eta,v)$ as
%%%
\begin{equation}
\setlength{\nulldelimiterspace}{0pt}
\left\{
\begin{IEEEeqnarraybox}[\relax][c]{l's}
x=u \cos kv-\eta \sin kv,\\
y=\eta \cos kv + u \sin kv,\\
z=v,
\end{IEEEeqnarraybox}
\right.
\label{polar-helicoid}
\end{equation}
%%%
where $k=2\pi/L_{\rm p}$ is the wavenumber of the helicoid and $L_{\rm p}$ is the twist pitch length.
The tape surface corresponds to $-w_0/2 < u < w_0/2$, $\eta=0$, and $-\infty< v  < \infty$.
The current density vector is expressed as $\bm{J}=\bm{\nabla}\times \bm{T}$ with the current vector potential $\bm{T}$
because of $\bm{\nabla}\cdot\bm{J}=0$.
In the thin-film limit (i.e., a strip with an infinitesimal thickness),
the electric current is restricted within the tape surface, $\eta=0$,
and thus one may use $\bm{T}=T\hat{\bm{n}}$ \cite{zhang2017},
where $\hat{\bm{n}}$ is the local unit vector normal to the tape surface and given on a twisted tape surface by
%It is thus convenient to introduce the current vector potential $\bm{T}$ \cite{zhang2017}:
%%%
\begin{eqnarray}
\hat{\bm{n}}
=\hat{\bm{v}}\times\hat{\bm{u}}
=\frac{\bm{\nabla}\eta}{|\bm{\nabla}\eta|}
=\frac{\bm{\nabla}\eta}{\sqrt{1+k^2u^2}},
\end{eqnarray}
%%%%
where $\hat{\bm{u}}$ and $\hat{\bm{v}}$ are the unit vectors along the $u$ and $v$ axes, respectively (see figure \ref{fig1}).
%Here, $\bm{\nabla}_{\rm T}$ denotes the spatial differential operator along the tangential direction of the tape surface.
The use of $\bm{T}=g(u,v)\bm{\nabla}\eta$ with $g(u,v)\equiv T(u,v)/\sqrt{1+k^2u^2}$
simplifies the analytic calculation of the current density on a twisted strip because of $\bm{\nabla}\times\bm{\nabla}\eta=0$.
%as $\bm{J}(u,v)=\bm{\nabla}g(u,v)\times \bm{\nabla}\eta$.
The current density at the tape surface, $\eta=0$, is thus obtained as 
%%%%
\begin{eqnarray}
\bm{J}(u,v)
=%\bm{\nabla}\times \left[T(u,v)\hat{\bm{n}} \right]
\bm{\nabla}g(u,v)\times \bm{\nabla}\eta
\label{current-density}
=J_u\hat{\bm{u}}+J_v\hat{\bm{v}},
\label{current}
\end{eqnarray}
%%%%
with
%%%
\begin{eqnarray}
J_u=-\frac{\partial g}{\partial v},~~J_v=\frac{\partial g}{\partial u}\sqrt{1+k^2u^2}.
\label{current-density-components}
\end{eqnarray}
%%%
Note that equation~(\ref{current-density}) satisfies both constraint conditions for the current density,
$\bm{J}\cdot\bm{\nabla}\eta=0$ and $\bm{\nabla}\cdot\bm{J}=0$.

%%%%%%%%%%%%%%%%%%%%%%%%%%%%%%%%%%%%%%%%%%%%%%%%%%%%%%%%%%%%%%%%%%%%
%%%%%%%%%%%%%%%%%%%%%%%%%%%%%%%%%%%%%%%%%%%%%%%%%%%%%%%%%%%%%%%%%%%%
%\section{Electromagnetic response on a twisted superconducting tape wire with constantly ramping fields}
\section{Electromagnetic response on a twisted superconducting tape with constantly ramping fields}
%%%%%%%%%%%%%%%%%%%%%%%%%%%%%%%%%%%%%%%%%%%%%%%%%%%%%%%%%%%%%%%%%%%%
%%%%%%%%%%%%%%%%%%%%%%%%%%%%%%%%%%%%%%%%%%%%%%%%%%%%%%%%%%%%%%%%%%%%
In a practical MRI magnet,
a transport current flows along SC wires,
%coils,
thereby generating a high external magnetic field.
Because the MRI magnet operates at high magnetic fields,
we consider the effect of the external field on the SC tape %tape wire
to be much more crucial than that of the transport current. 
We then start with Faraday's law in steadily ramped magnetic fields as in figure~\ref{fig3},
%%%%%
\begin{eqnarray}
\bm{\nabla} \times \bm{E}=-\frac{\partial \bm{B}}{\partial t}\approx -\beta \hat{\bm{y}},
\label{Faraday}
\end{eqnarray}
%%%%%
where $\beta$ denotes the sweep rate of the magnetic field.
$\hat{\bm{y}}$ is the unit vector in the $y$ direction.
Throughout this paper, the sweep rate is fixed at $\beta=4.69$
%$\beta=4.6875$
mT/s unless otherwise specified.
The magnetic field due to the screening current
in the right-hand side of equation~(\ref{Faraday}) is neglected.
We consider the response to the magnetic field component perpendicular to the tape surface
because the response to the parallel field component can be neglected in the thin-film limit.
Thus, taking the projection of equation~(\ref{Faraday}) onto $\bm{\nabla}\eta$ affords
%%%%
\begin{eqnarray}
\left( \bm{\nabla}\times\bm{E} \right)\cdot\bm{\nabla}\eta
=
-\beta\hat{\bm{y}}\cdot \bm{\nabla}\eta
=-\beta \cos kv.
\label{Faraday-projection}
\end{eqnarray}
%%%%
Using the electric field in the coordinate system of an SC helicoid as described in equation (\ref{polar-helicoid}), 
equation (\ref{Faraday-projection}) on the tape surface, $\eta=0$, reduces to
%%%%
\begin{eqnarray}
\frac{\partial}{\partial u}\left[ E_v\sqrt{1+k^2u^2}\right]-\frac{\partial E_u}{\partial v}=\beta \cos kv.
\label{Faraday_electric-field}
\end{eqnarray}
%%%%
Identifying the scalar function $g(u,v)$, which determines the current flow lines, is necessary for performing the analytic calculation of magnetization and loss power.
The electric field $\bm{E}$--current density $\bm{J}$ characteristics of the rare earth--barium--copper oxide  SC tape %tape wire
are assumed to be described by the power law,
%%%
\begin{align}
\bm{E}
&=\rho_{\rm sc}(|\bm{J}|)\bm{J},
\label{e-j}
\\
\rho_{\rm sc}(|\bm{J}|)
&=\frac{E_{\rm c}}{J_{\rm c}}\left( \frac{|\bm{J}|}{J_{\rm c}} \right)^{n-1},
\label{n-value model}
\end{align}
%%%
where isotropic SC nonlinear resistivity is assumed.
$E_{\rm c}$ is the electric field criterion.
From equations (\ref{current-density-components}) and (\ref{e-j}),
we obtain the electric field on the tape surface,
%%%%
\begin{eqnarray}
E_u=-\rho_{\rm sc}\frac{\partial g}{\partial v},~~E_v=\rho_{\rm sc}\frac{\partial g}{\partial u}\sqrt{1+k^2u^2}.
\label{electric-field}
\end{eqnarray}
%%%%
By substituting equation (\ref{electric-field}) into equation (\ref{Faraday_electric-field}),
the equation for $g(u,v)$ on the tape surface is obtained as
%%%%
\begin{eqnarray}
\frac{\partial}{\partial u}\left[ \rho_{\rm sc}\frac{\partial g}{\partial u}(1+k^2u^2) \right]+\frac{\partial}{\partial v}\left( \rho_{\rm sc}\frac{\partial g}{\partial v}\right)
=
\beta \cos kv.
\label{Faraday_current}
\end{eqnarray}
%%%%

%%%%%%%%%%%%%%%%%%%%%%%%%%%%%%%%%%%%%%%%%%%%%%%%%%%%%%%%%%%%%%%%%%
%%%%%%%%%%%%%%%%%%%%%%%%%%%%%%%%%%%%%%%%%%%%%%%%%%%%%%%%%%%%%%%%%%
\section{Analysis in the loosely twisted tape limit}
%%%%%%%%%%%%%%%%%%%%%%%%%%%%%%%%%%%%%%%%%%%%%%%%%%%%%%%%%%%%%%%%%%
%%%%%%%%%%%%%%%%%%%%%%%%%%%%%%%%%%%%%%%%%%%%%%%%%%%%%%%%%%%%%%%%%%
Here, we analytically derive the formulae for both magnetization and loss power
on the basis of equation~(\ref{Faraday_current})
in the loosely twisted tape limit of $kw_0\rightarrow 0$ (i.e., $L_{\rm p}\rightarrow \infty$).
A loosely twisted stacked-tape cable conductor %wire
with $w_0/L_{\rm p}\sim 0.01$ has indeed been fabricated using the currently available technology \cite{takayasu2012-aip}.

By taking the limit of $k\rightarrow0$ and $v\rightarrow \infty$ and keeping $\phi=kv$ over a twist pitch on a tape, %tape wire,
%greater than the twist pitch on a tape wire, 
equation~(\ref{Faraday_current}) reduces to
%%%%
\begin{eqnarray}
\frac{\partial}{\partial u} \left(\rho_{\rm sc} \frac{\partial g}{\partial u} \right) 
=\beta \cos\phi.
\label{zero-th_Maxwell}
\end{eqnarray}
%%%%
As for the SC nonlinear resistivity $\rho_{\rm sc}$,
by carrying out the perturbation expansion with respect to $kw_0$
and taking the limit of $k\rightarrow 0$,
we obtain
%%%%
$
|\bm{J}|^{n-1}
=| \partial g/ \partial u|^{n-1}.
$
%%%%
Consequently, the SC nonlinear resistivity becomes
%%%%
\begin{eqnarray}
\rho_{\rm sc}=\frac{E_{\rm c}}{J^n_{\rm c}}\left|{\partial g\over \partial u}\right|^{n-1}.
\label{zero-th_rho}
\end{eqnarray}
%%%%
By substituting equation~(\ref{zero-th_rho}) into equation~(\ref{zero-th_Maxwell}),
the solution of the Maxwell equation in the loosely twisted tape limit is obtained as
%%%%
\begin{eqnarray}
g(u,\phi)=\frac{{\rm sgn}(\cos\phi)|u|J_{\rm c}}{1+1/n}\left( \frac{\beta |u\cos\phi|}{E_{\rm c}} \right)^{1/n}+C(\phi),
\end{eqnarray}
%%%%%
where the integral function, $C(\phi)$, is determined to satisfy the Dirichlet boundary condition at the edges of the long dimension, $g(u=\pm w_0/2,\phi)=0$, as follows:
%%%%%
\begin{eqnarray}
C(\phi)=-\frac{{\rm sgn}(\cos \phi)w_0J_{\rm c}}{2(1+1/n)}\left( \frac{\beta w_0}{2E_{\rm c}}\right)^{1/n}|\cos \phi|^{1/n}.
\end{eqnarray}
%%%%%

Next, in the coordinate system of the SC helicoid [equation~(\ref{polar-helicoid})],
an arbitrary vector
%physical quantity
$\bm{O}=(O_x,O_y,O_z)$ (e.g., $\bm{O}=\bm{E}, \bm{J}$) in the loosely twisted tape limit ($kw_0\rightarrow 0$) is expressed as
%%%%
\begin{equation}
\setlength{\nulldelimiterspace}{0pt}
\left\{
\begin{IEEEeqnarraybox}[\relax][c]{l's}
O_x=O_u \cos \phi-O_\eta \sin \phi,\\
O_y=O_u \sin \phi+O_\eta \cos \phi,\\
O_z=O_v.
\end{IEEEeqnarraybox}
\right.
%\label{polar-helicoid}
\end{equation}
%%%%
Hence, the component of magnetization parallel to ${\bm B}_{\rm a}$ ($||~\hat{\bm y}$) in the loosely twisted tape limit can be evaluated as
%%%%
\begin{align}
M_y
&=\frac{1}{w_0L_{\rm p}} \int_V{\rm d}V(zJ_x-xJ_z)\nonumber\\
&\approx \frac{d_0}{w_0 L_{\rm p}}\int_{-w_0/2}^{w_0/2}{\rm d}u \int_0^{L_{\rm p}}{\rm d}v(vJ_u \cos \phi-u \cos \phi J_v),\nonumber\\
\end{align}
%%%%
with $J_u=-\partial g/\partial v$ and $J_v=\partial g/\partial u$ [equation (\ref{current-density-components}) with $k \rightarrow 0$].
The $J_\eta$ term disappears because the current density component perpendicular to the tape surface ($\eta=0$) vanishes in the sheet current approximation.
We further neglect $J_u$ because $J_u=-k(\partial g/\partial \phi)$ vanishes in the loosely twisted tape limit ($kw_0\rightarrow 0$).
Hence,
%%%%
\begin{align}
M_y&\approx-\frac{d_0}{w_0 L_{\rm p}}\int_{-w_0/2}^{w_0/2}{\rm d}u \int_0^{L_{\rm p}}{\rm d}v
u \cos \phi J_v\nonumber \\
&=-\frac{B({2n+1\over 2n},{1\over 2})}{\pi}
\left( \beta w_0 \over 2E_{\rm c}\right)^{1/n}\frac{J_{\rm c}w_0 d_0}{2(2+1/n)},
\label{magnetization}
\end{align}
%%%%
where $B(p,q)=2\int_0^{\pi/2}{\rm d}\theta \cos^{2p-1}\theta \sin^{2q-1}\theta$ is the beta function
with the positive real numbers $p$ and $q$.
The loss per unit length can be similarly evaluated from $J_v$ and $E_v=\rho_{\rm sc}(\partial g/\partial u)$ as
%%%%
\begin{align}
Q
&=\frac{1}{L_{\rm p}}\int_V {\rm d}V(E_uJ_u+E_vJ_v)\nonumber\\
&\approx
\frac{d_0}{L_{\rm p}}\int_0^{L_{\rm p}}{\rm d}v  \int_{-w_0/2}^{w_0/2}{\rm d}u E_vJ_v\nonumber\\
&=
\frac{B(\frac{2n+1}{2n},\frac{1}{2})}{\pi}\left( \frac{\beta w_0}{2E_{\rm c}}\right)^{1+1/n}\frac{E_{\rm c}J_{\rm c}w_0 d_0}{2+1/n}.
\label{loss-theory-loosely-twisted-tape-limit}
\end{align}
%%%%

It should be noted that in the loosely twisted tape limit of $L_{\rm p}\rightarrow \infty$,
both magnetization and loss per unit length are independent of $L_{\rm p}$ 
and become smaller than those for a flat tape by a factor of $B(1+1/2n,1/2)/\pi$ owing to twisting alone.
Because this factor reduces to the geometric factor $2/\pi$, which is the area of an SC helicoid projected onto the $x$--$z$ plane in figure \ref{fig1}, in the Bean limit of $n\rightarrow \infty$ \cite{takayasu2012,grilli2015},
it can be understood that the area of the tape %tape wire
interlinking across the magnetic fluxes decreases upon twisting.

%%%%%%%%%%%%%%%%%%%%%%%%%%%%%%%%%%%%%%%%%%%%%%%%%%%%%%%%%%%%%%%%%%%%
%%%%%%%%%%%%%%%%%%%%%%%%%%%%%%%%%%%%%%%%%%%%%%%%%%%%%%%%%%%%%%%%%%%%
\section{Tilted flat tape approximation}
%%%%%%%%%%%%%%%%%%%%%%%%%%%%%%%%%%%%%%%%%%%%%%%%%%%%%%%%%%%%%%%%%%%%
%%%%%%%%%%%%%%%%%%%%%%%%%%%%%%%%%%%%%%%%%%%%%%%%%%%%%%%%%%%%%%%%%%%%
In this section,
we intuitively explain how the formula of magnetization obtained from the tilted flat tape approximation, which is originally mentioned for the SC tapes helically wound on a former in Ref. \cite{souc2010}, coincides with that obtained from
the solution of equation~(\ref{Faraday_current}) in the loosely twisted tape limit.
The precedent studies on twisted stacked tape cables in the presence of a transport current have numerically demonstrated the validity of the tilted flat tape view of a twisted tape by looking into loss power density of the twisted tape \cite{grilli2015,kruger2015}. 
%and show
%a tilted flat tape
We also show that
%the combination
the sum of a portion of tilted flat tapes of infinite length 
can serve as a good approximation for a twisted SC tape %tape wire
when the twist pitch length is sufficiently high relative to the tape width ($w_0 \ll L_{\rm p}$).

%%%%%%%%%%%%%%%%%%%%%%%%%%%%%%%%%
\subsection{Perpendicular field distribution}
%%%%%%%%%%%%%%%%%%%%%%%%%%%%%%%%%
%%%%%%%%%%%%%%%%%%%%%%%%%%%%%%%%%
%%%%%%%%%%%%%%%%%%%%%%%%%%%%%%%%%
\begin{figure}[tb]
  \begin{center}
    \begin{tabular}{p{85mm}}
      \resizebox{85mm}{!}{\includegraphics{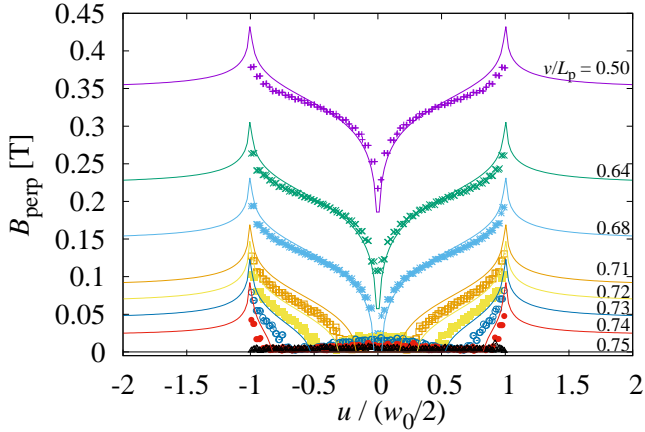}}
    \end{tabular}
\caption{
\label{fig2}
Spatial profiles of the perpendicular magnetic field component on a twisted tape %tape wire
for the applied field $B_{\rm a}=0.35$ T and $L_{\rm p}=40$ mm.
The symbols and solid lines indicate the numerical and analytic results, respectively.
The position on a twisted strip $v/L_{\rm p}$ is indicated for each curve.
}
  \end{center}
\end{figure}
%%%%%%%%%%%%%%%%%%%%%%%%%%%%%%%%%
%%%%%%%%%%%%%%%%%%%%%%%%%%%%%%%%%
In figure~\ref{fig2},
the symbols represent the numerical results for the perpendicular magnetic field distribution $B_{\rm perp}(u)$ on a twisted SC tape %tape wire
upon changing the surface parameter $v$.
The detailed numerical results are described in section \ref{numerical-results}.
The solid lines represent the analytic solutions obtained by Brandt and Indenbom \cite{brandt1993}
using the applied-field component perpendicular to the tape surface,
%$B_{{\rm a}\perp}(v)=|B_{\rm a}\cos(2\pi v/L_{\rm p})|$, with the applied field $B_{\rm a}=0.35$ T.
$|B_{{\rm a}\eta}(v)|=|B_{\rm a}\cos(2\pi v/L_{\rm p})|$, with the applied field $B_{\rm a}=0.35$ T.
The numerical results for $v/L_{\rm p}=0.50$--$0.68$ are well fitted by the analytic solutions,
and the characteristics of $B_{\rm perp}$ for $v/L_{\rm p}=0.71$--$0.75$ are also roughly reproduced by the analytic solutions.
On the basis of the numerical results shown in figure~\ref{fig2}, the tilted flat tape approximation can be considered a good approximation for analyzing a twisted tape %tape wire.
%We will analytically verify that in the next section. 

%%%%%%%%%%%%%%%%%%%%%%%%%%%%%%%%%
\subsection{Tilted flat tape approximation for magnetization}
%%%%%%%%%%%%%%%%%%%%%%%%%%%%%%%%%
We next consider an external magnetic field swept at a constant rate $\beta$
and neglect the magnetic field due to the screening current.
%self-field due to the screening current.
The magnetization component parallel to the external magnetic field, ${\rm d}\bm{B}_{\rm a}/{\rm d}t=\beta \hat{\bm{y}}$,
in the constantly swept time domain contributes to the loss power.

First, let $M_\eta(B_{\rm a}\cos kv)$ be the magnetization component normal to the tape surface
in response to the field component perpendicular to the tape surface,
$B_{{\rm a}\eta}(v)=B_{\rm a}\cos kv$.
%$B_{\rm a}\cos kv$.
Taking the projection of $M_\eta$ onto the applied field direction ($||~\hat{\bm{y}}$), $M_\eta(B_{\rm a}\cos kv)\cos kv$, and 
integrating it over a twist pitch,
the magnetization component parallel to $\bm{B}_{\rm a}$ is evaluated from
%%%%
\begin{align}
M_y&=\frac{d_0}{L_{\rm p}}\int_0^{L_{\rm p}}{\rm d}vM_\eta \left(B_{\rm a}\cos kv\right)\cos kv,\\
M_\eta(v)&\equiv \frac{1}{w_0}\int_{-w_0/2}^{w_0/2}{\rm d}u (vJ_u-uJ_v) \nonumber\\
&\approx -\frac{1}{w_0}\int_{-w_0/2}^{w_0/2}{\rm d}u uJ_v,
\end{align}
%%%%
%Here the orthogonal coordinate $(r_\parallel,r_\perp,z)$ is used for the tilted flat tape.
%Here $M_\eta$ is the magnetization component normal to the tape surface.
where $J_u$ is assumed to be vanishingly small in the tilted flat tape approximation and is neglected here.
In the case of a tilted flat tape of infinite length,
because there is no current winding, that is, the current component in the width direction is absent ($J_u=0$),
$E_u =0$ can be deduced and $\partial E_u/\partial v$ can be dropped from equation (\ref{Faraday_electric-field}).
Thus, for a tilted flat tape of infinite length in a steady state,
we obtain $E_v\approx u\beta\cos kv$ \cite{mawatari1997}
by taking the limit of $k\rightarrow0$ and $v\rightarrow \infty$
and keeping $\phi=kv$ at the right-hand side of equation (\ref{Faraday_electric-field}).
With an electric field of $E_v$, the current density in the direction of the long dimension can be calculated as $J_v=J_{\rm c}(|E_v|/E_{\rm c})^{-1+1/n} E_v/E_{\rm c}$
from equations (\ref{e-j}) and (\ref{n-value model}).
%For the perpendicular component of the constantly ramping magnetic field $\beta \cos kz \hat{\bm{r}}_\perp$,
%$E_z\approx r_\parallel \beta \cos kz$ is obtained \cite{mawatari1997}.
%It is plausible to neglect the parallel electric-field component $E_\parallel$ in the case of $w_0\ll L_{\rm p}$.
Hence, the magnetization component parallel to $\bm{B}_{\rm a}$ exactly coincides with equation (\ref{magnetization}).
This is plausible because the correction terms due to the twisting, which is dependent on $k$, vanish
in the tilted infinite-length flat tape approximation,
in which $L_{\rm p}\rightarrow \infty$.
%when $L_{\rm p}\rightarrow \infty$.
%in the Maxwell equation in the limit of $L_{\rm p}\rightarrow \infty$.

%%%%%%%%%%%%%%%%%%%%%%%%%%%%%%%%%%%%%%%%%%%%%%%%%%%%%%%%%%%%%%%%%%%%
%%%%%%%%%%%%%%%%%%%%%%%%%%%%%%%%%%%%%%%%%%%%%%%%%%%%%%%%%%%%%%%%%%%%
\section{Numerical results}
\label{numerical-results}
%%%%%%%%%%%%%%%%%%%%%%%%%%%%%%%%%%%%%%%%%%%%%%%%%%%%%%%%%%%%%%%%%%%%
%%%%%%%%%%%%%%%%%%%%%%%%%%%%%%%%%%%%%%%%%%%%%%%%%%%%%%%%%%%%%%%%%%%%
%%%%%%%%%%%%%%%%%%%%%%%%%%%%%%%%%
\subsection{Temporal loss power and twist-pitch dependence}
\label{Temporal loss power and twist pitch dependence}
%%%%%%%%%%%%%%%%%%%%%%%%%%%%%%%%%
%%%%%%%%%%%%%%%%%%%%%%%%%%%%%%%%%
%%%%%%%%%%%%%%%%%%%%%%%%%%%%%%%%%
\begin{figure}[tb]
  \begin{center}
    \begin{tabular}{p{85mm}}
      \resizebox{85mm}{!}{\includegraphics{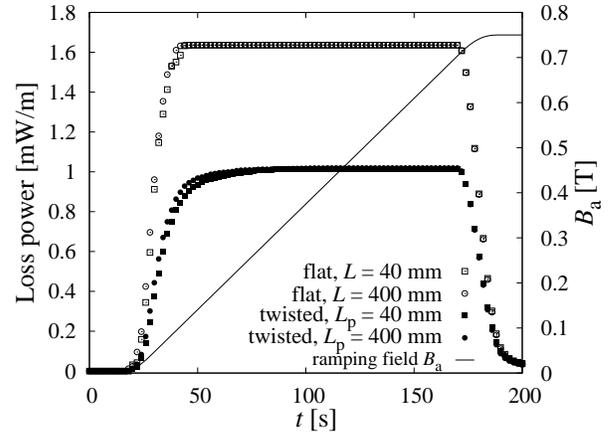}}
    \end{tabular}
\caption{
\label{fig3}
Temporal loss power per unit length for flat tapes with a wire length of $L=40$ or $400$ mm and twisted tapes with a twist pitch length of $L_{\rm p}=40$ or $400$ mm.
The solid line indicates the ramped applied field. 
}
  \end{center}
\end{figure}
%%%%%%%%%%%%%%%%%%%%%%%%%%%%%%%%%
%%%%%%%%%%%%%%%%%%%%%%%%%%%%%%%%%
%%%%%%%%%%%%%%%%%%%%%%%%%%%%%%%%%
%%%%%%%%%%%%%%%%%%%%%%%%%%%%%%%%%
\begin{figure}[tb]
  \begin{center}
    \begin{tabular}{p{80mm}}
      \resizebox{80mm}{!}{\includegraphics{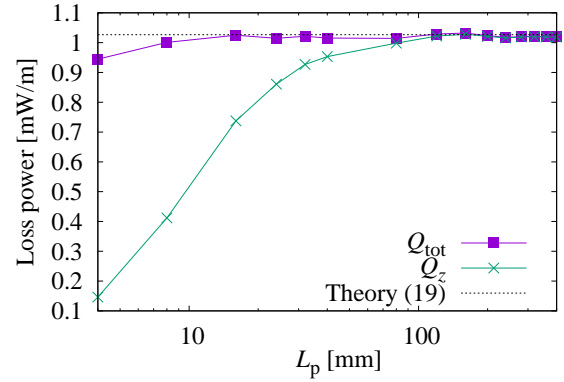}}
    \end{tabular}
\caption{
\label{fig4}
Twist-pitch dependences of the total loss power per unit length (squares) and the contributions from $J_z$ and $E_z$ to the loss power per unit length (crosses).
The dotted line represents the theoretical value obtained using equation~(\ref{loss-theory-loosely-twisted-tape-limit}).
}
  \end{center}
\end{figure}
%%%%%%%%%%%%%%%%%%%%%%%%%%%%%%%%%
%%%%%%%%%%%%%%%%%%%%%%%%%%%%%%%%%
In the numerical calculations,
the following equation, which is equivalent to equation (\ref{Faraday-projection}) owing to the identity $\bm{\nabla}\times\bm{\nabla}\eta=0$, was solved utilizing the commercial software COMSOL \cite{comsol}: 
%%%%
\begin{eqnarray}
\bm{\nabla}\cdot\left( \bm{E} \times \bm{\nabla}\eta \right)
=
-\frac{\partial \bm{B}}{\partial t}\cdot \bm{\nabla}\eta.
\end{eqnarray}
%%%% 
The periodic boundary condition was imposed in the $z$ direction.
The parameters of the SC tape %tape wire
were set to
$w_0=4$ mm, $d_0=2$ $\mu$m, $J_{\rm c}=5\times 10^{10}$ A/m$^2$, $E_{\rm c}=1$ $\mu$V/cm, and $n=21$.
Figure~\ref{fig3} presents the numerical results for the temporal loss power per unit length for both flat and twisted tapes. %tape wires.
For a twisted tape, the value of loss power per unit length (filled squares and circles)
becomes saturated toward $Q_{\rm twist}\approx 1.0$ mW/m even after changing the twist pitch length $L_{\rm p}$, showing that it is not dependent on $L_{\rm p}$
when the condition $w_0\ll L_{\rm p}$ is satisfied.
Owing to the twisting alone, the loss power value is reduced from that in the flat tape %tape wire
(open squares and circles), $Q_{\rm flat}\approx 1.636$ mW/m.
For $L_{\rm p}=400$ mm, $Q_{\rm twist}/Q_{\rm flat}\approx 0.623$, which shows good agreement with the theoretically evaluated value from equation~(\ref{loss-theory-loosely-twisted-tape-limit}),
$B({1+1/2n},{1/2})/\pi\approx 0.628$ for $n=21$.

In figure~\ref{fig4}, we show the dependence of the loss power per unit length on $L_{\rm p}$.
We define the value of loss power in the constantly ramping field as that at $t = 170$ s in figure~\ref{fig3}.
The dotted line indicates the theoretically evaluated value based on equation~(\ref{loss-theory-loosely-twisted-tape-limit}).
The numerical values of $Q_{\rm tot}$ (symbols) agree well with the theoretical value, with a deviation of less than $2\%$ for $L_{\rm p} > 2w_0$,
%$\sim 1\%$ for $w_0\ll L_{\rm p}$,
and $Q_{\rm tot}$ is independent of $L_{\rm p}$.
Here, $Q_{\rm tot}$ is defined by $Q_{\rm tot}=(d_0/L_{\rm p})\int_S {\rm d}S \bm{E}\cdot \bm{J}$.
$Q_z$ shows the contributions from $J_z$ and $E_z$ to the loss power, although these are not observable, 
and it dominates the loss power for $L_{\rm p}\gg w_0$, saturating toward the theoretical value.
The noticeably different behavior between $Q_{\rm tot}$ and $Q_z$ with decreasing $L_{\rm p}$ stems from the winding of the current flow on a strip.

%%%%%%%%%%%%%%%%%%%%%%%%%%%%%%%%%%%%%%%%%%%%%%%%%%%%%%%%%%%%%%%%%%%%
\subsection{Effect of field and its angle-dependent critical current density}
%%%%%%%%%%%%%%%%%%%%%%%%%%%%%%%%%%%%%%%%%%%%%%%%%%%%%%%%%%%%%%%%%%%%
%%%%%%%%%%%%%%%%%%%%%%%%%%%%%%%%%
%%%%%%%%%%%%%%%%%%%%%%%%%%%%%%%%%
\begin{figure}[tb]
  \begin{center}
    \begin{tabular}{p{85mm}}
      \resizebox{85mm}{!}{\includegraphics{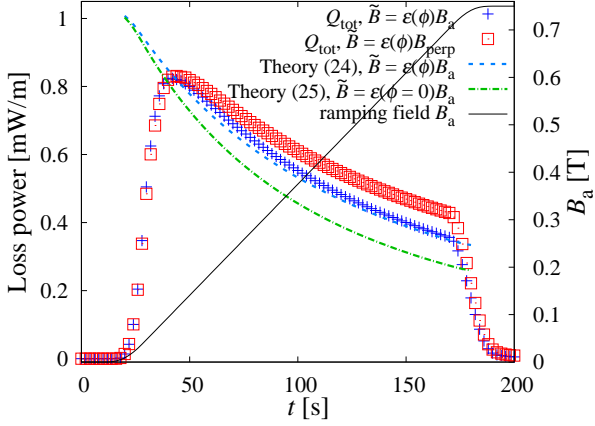}}
    \end{tabular}
\caption{
\label{fig5}
Temporal loss power per unit length taking into account the field and/or field-angle dependence of $J_{\rm c}$ for $L_{\rm p}=50w_0$.
The symbols represent the numerical results.
The dashed and dash-dot lines represent the theoretical results evaluated via equation (\ref{loss-theory-integrated}) and equation (\ref{loss-theory-phi0}), respectively.
The solid line shows the ramped applied field.
}
  \end{center}
\end{figure}
%%%%%%%%%%%%%%%%%%%%%%%%%%%%%%%%%
%%%%%%%%%%%%%%%%%%%%%%%%%%%%%%%%%
Next, the effect of the field and its angle-dependent critical current density on the temporal loss power is discussed.
Regarding the angular dependence,
%In terms of the angular dependence,
we show that the loss power evaluated with $J_{\rm c}$
in an applied magnetic field perpendicular to the tape surface (i.e., parallel to the $c$ axis)
%in a perpendicular field parallel to the c-axis
can serve as
an approximate estimate
%a rough estimate
of the loss power value.
We adopt the generalized Kim model as a simple model for the field and its angle dependences of $J_{\rm c}$ \cite{kim1962,pardo2011}:
%%%
\begin{eqnarray}
J_{\rm c}(B,\phi)=\frac{J_{{\rm c}0}}{\left( 1+\tilde{B}/B_0\right)^\alpha},
\end{eqnarray}
%%%
where $J_{{\rm c}0}$ is the critical current density in the absence of the magnetic field and $B_0$ and $\alpha$ are constants.
The uniaxial anisotropy of SC materials with isotropic (randomly distributed) disorders can be incorporated into this theory through the temporal effective local magnetic field
$\tilde{B}=\varepsilon(\phi)B$,
where $B$ is the temporal local field, $\varepsilon(\phi)=\sqrt{\cos^2\phi+\gamma^{-2}\sin^2\phi}$ is the scaling factor, and the angle $\phi=kv$ is measured from the $c$ axis \cite{blatter1992,civale2004-apl,civale2004}. 
$\gamma=\sqrt{m_{\rm c}/m_{\rm ab}}$ is the mass anisotropy ratio.
The masses $m_{\rm ab}$ and $m_{\rm c}$ characterize the transport in the SC state within the $a$--$b$ plane and in the direction of the $c$ axis, respectively.
The mass anisotropy ratio is set to $\gamma=5$ assuming a SC YBa$_2$Cu$_3$O$_7$ thin film \cite{civale2004-apl}. 
The constants $B_0=0.75$ T and $\alpha=2$ are fixed throughout the paper, and the other parameters are the same as those used in subsection~\ref{Temporal loss power and twist pitch dependence}. 

In figure \ref{fig5}, the symbols represent the numerical results of the temporal loss power for $L_{\rm p}=50w_0$.
The square and plus symbols indicate the numerical results for $\tilde{B}=\varepsilon(\phi)B_{\rm perp}$ and $\tilde{B}=\varepsilon(\phi)B_{\rm a}$, respectively.
Here, $B_{\rm perp}$ denotes the temporal local perpendicular magnetic field, through which
the contribution to the loss power from the magnetic field due to the screening current is incorporated in the numerical simulation.
%the self-field contribution to the loss power is incorporated in the numerical simulation.
The choice of $\tilde{B}=\varepsilon(\phi)B_{\rm perp}$ affords the most accurate numerical results in the present study. 
In contrast, the contribution from the magnetic field due to the screening current
partially lacks
%is somewhat underrepresented
when $\tilde{B}=\varepsilon(\phi)B_{\rm a}$ is adopted,
although this approximation is plausible because the magnetic field due to the screening current
on a strip is small.

The loss power decreases due to the degradation of $J_{\rm c}$
with constant ramping of $B_{\rm a}$.
%with continuous ramping of $B_{\rm a}$.
The slight increase in the loss power upon adopting $\tilde{B}=\varepsilon(\phi)B_{\rm perp}$ reflects
the contribution from the magnetic field due to the screening current.
%the self-field contribution.
The analytic evaluation of the temporal loss power (dashed line in figure~\ref{fig5}), 
%%%
\begin{align}
Q(t) =&\left( \frac{\beta w_0}{2E_{\rm c}}\right)^{1+1/n}\frac{J_{{\rm c}0}E_{\rm c}d_0w_0}{2+1/n}\nonumber \\
&\times \frac{2}{\pi} \int_0^{\pi/2}{\rm d}\phi
\frac{\cos^{1+1/n}\phi}{\left[ 1+\varepsilon(\phi)B_{\rm a}(t)/B_0 \right]^2}
\label{loss-theory-integrated}
\end{align}
%%%
agrees with the numerical results (plus symbols in figure~\ref{fig5}). 
%The factor dependent on $k$ in $J_{\rm c}(\tilde{B})$ is kept over the one twist pitch length $L_{\rm p}$ to obtain equation~(\ref{loss-theory-integrated}).
The first line in equation~(\ref{loss-theory-integrated}) corresponds to the loss power in a flat tape of infinite length.
The factor $2/\pi$ is due to twisting, and the $\phi$ integral arises from both the field-angle dependence of $J_{\rm c}$ and twisting.
Note that the theoretical curves in figure~\ref{fig5} are valid only while the magnetic field is constantly ramped (solid line in figure~\ref{fig5} for $20\le t\le180$),
because we assume the constant sweep rate $\beta$.

Intuitively, it is expected that the loss power is not generated so much from the portion of the tape surface parallel to $\bm{B}_{\rm a}(t)=B_{\rm a}(t)\hat{\bm{y}}$,
because the tape surface does not magnetize in the field direction ($||~\hat{\bm{y}}$).
Thus, it might serve as a good approximation to neglect the field-angle dependence of $J_{\rm c}$ and let $J_{\rm c}(B,\phi)$
be the field-angle-independent value $J_{\rm c}(B_{\rm a}, \phi=0)$
in the applied field perpendicular to the tape surface.
In this case, $\varepsilon(\phi=0)=1$, and the temporal loss power can therefore be recast as
%%%
\begin{align}
Q(t) =&\left( \frac{\beta w_0}{2E_{\rm c}}\right)^{1+1/n}\frac{J_{{\rm c}0}E_{\rm c}d_0w_0}{2+1/n}\nonumber\\
&\times \frac{B({2n+1\over 2n},{1\over2})}{\pi} \frac{1}{\left[ 1+B_{\rm a}(t)/B_0 \right]^2}
\label{loss-theory-phi0}
\end{align}
%%%
Because the field-angle dependence of $J_{\rm c}$ is dropped,
equation~(\ref{loss-theory-phi0}) (dash-dot line in figure~\ref{fig5}) slightly underestimates the numerically evaluated temporal loss power including
the contribution from the magnetic field due to the screening current (squares in figure~\ref{fig5}),
%the self-field contribution (squares in figure~\ref{fig5}),
although it works sufficiently well to provide a rough numerical estimate of the loss power value.

%%%%%%%%%%%%%%%%%%%%%%%%%%%%%%%%%%%%%%%%%%%%%%%%%%%%%%%%%%%%%%%%%%%%
\subsection{Sweep-rate dependence}
%%%%%%%%%%%%%%%%%%%%%%%%%%%%%%%%%%%%%%%%%%%%%%%%%%%%%%%%%%%%%%%%%%%%
%%%%%%%%%%%%%%%%%%%%%%%%%%%%%%%%%
%%%%%%%%%%%%%%%%%%%%%%%%%%%%%%%%%
\begin{figure}[tb]
  \begin{center}
    \begin{tabular}{p{80mm}}
      \resizebox{80mm}{!}{\includegraphics{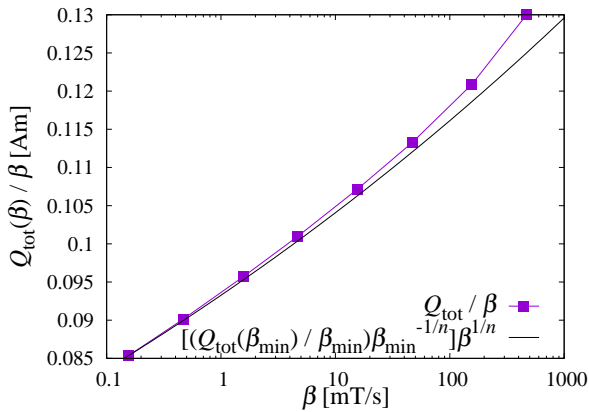}}
    \end{tabular}
\caption{
\label{fig6}
Field sweep-rate $\beta$ dependence of the loss power relative to $\beta$ for $L_{\rm p}=50w_0$. 
The closed squares and the solid line  indicate the numerical results and the theoretical curve, respectively.
}
  \end{center}
\end{figure}
%%%%%%%%%%%%%%%%%%%%%%%%%%%%%%%%%
%%%%%%%%%%%%%%%%%%%%%%%%%%%%%%%%%
The steady-state assumption, $\partial B/\partial t=\beta$, might fail as a result of the dependence of $J_{\rm c}$ on the field and its angle. 
Also, for large field sweep rates $\beta$,
the situation might deviate from the assumptions used to derive the analytic formula for the loss power [equation (\ref{loss-theory-integrated})],
that is,
the steady state of the constantly ramped magnetic field and absence of a magnetic field due to the screening current.
Therefore, the $\beta$ dependence of the loss power is discussed to explore the behavior of the loss power at large values of $\beta$.

Figure~\ref{fig6} shows the $\beta$ dependence of the total loss power relative to $\beta$ 
in the steady state at $B_{\rm a}=0.6$ T, which is sufficiently higher than the field needed for full flux penetration $B_{\rm p}$.
We consider that $B_{\rm p}$ for a twisted tape can be approximately evaluated
via the formula for a flat tape of infinite length as 
$B_{\rm p}=(\mu_0J_{\rm c0}d_0/\pi)[1+\ln(w_0/d_0)]\approx0.344$ T \cite{brandt1996},
where $\mu_0$ is the magnetic permeability of a vacuum.
The solid line in figure~\ref{fig6} represents the theoretical curve [equation (\ref{loss-theory-integrated})] for the $\beta$ dependence of $Q/\beta$.
The vertical axis is rescaled for the theoretical curve so as to fit the numerical results (squares in figure~\ref{fig6})
by the power-law behavior $\beta^{1/n}$ expected from equation~(\ref{loss-theory-integrated}),
where $\beta_{\rm min}=0.156$ mT/s is the minimum sweep rate.
With increasing $\beta$, the numerical results for $Q_{\rm tot}/\beta$ gradually deviate from the $\beta^{1/n}$ behavior.
The deviation at large values of $\beta$ can be ascribed to the magnetic field due to the screening current
originated from the field and its angle-dependent $J_{\rm c}$,
%self-field,
%the screening-current induced field,
which is neglected in the analytic formula [equation~(\ref{loss-theory-integrated})].

%%%%%%%%%%%%%%%%%%%%%%%%%%%%%%%%%%%%%%%%%%%%%%%%%%%%%%%%%%%%%%%%%%%%
%%%%%%%%%%%%%%%%%%%%%%%%%%%%%%%%%%%%%%%%%%%%%%%%%%%%%%%%%%%%%%%%%%%%
\section{Summary and discussion}
%%%%%%%%%%%%%%%%%%%%%%%%%%%%%%%%%%%%%%%%%%%%%%%%%%%%%%%%%%%%%%%%%%%%
%%%%%%%%%%%%%%%%%%%%%%%%%%%%%%%%%%%%%%%%%%%%%%%%%%%%%%%%%%%%%%%%%%%%
The magnetization loss on a twisted SC strip in a constantly ramped magnetic field has been theoretically and numerically studied.
In the loosely twisted tape limit of $kw_0\rightarrow 0$ (i.e., $L_{\rm p}\rightarrow \infty$),
analytic formulae for both magnetization and loss power were derived on the basis of the equation for the scalar function $g$ in a helicoidal coordinate system.
The formula for the magnetization was rederived in an intuitive manner using the tilted flat tape approximation,
which was found to be a good approximation as the numerical value of loss power showed good agreement with the theoretical value obtained using this approximation.
% \rightarrow\infty$. 
Upon twisting an SC tape, %tape wire,
both magnetization and loss power per unit length decreased by a factor of $B(1+1/2n,1/2)/\pi$ compared with those in a flat tape,
but twisting alone had only a marginal effect on reducing these values.
%Besides, neither magnetization nor loss power per unit length depend on $L_{\rm p}$.
%With these knowledges,
Twisting a SC tape plays a crucial role when a SC tape is striated and coated by the normal metal stabilizer such as copper.
On the basis of the present approach,
we will discuss the condition that electromagnetic coupling is suppressed in a twisted multifilamentary SC tape %tape wire
in a swept magnetic field elsewhere.
The field-angle dependence of $J_{\rm c}$ due to the twist of an SC tape was considered to evaluate the loss power numerically,
and the results revealed that a rough estimate of the loss power is possible via $J_{\rm c}$ in a perpendicular field.

%The reason why the loss power is independent of $L_{\rm p}$ in the absence of multifilamentarization can be understood as follows.
%In the present study, the periodic boundary condition was imposed in the direction of $z$.
%Thus, the loss power was determined only via vortex penetration into a strip from the edges of the long dimension.
%The twist of a strip cannot change the length of the long dimension of the strip, and the loss power is therefore determined by the length of the short dimension of the strip, $Q\propto w^{2+1/n}_0$.

In more practical situations such as stacking or coiling of the SC tapes, %tape wires,
the effects of the self-field due to the transport current, the spatial inhomogeneity of $J_{\rm c}$, and the spatial inhomogeneity of the perpendicular component of the external magnetic field are crucial factors in the electromagnetic response.
The successful consideration of these factors will require the application of numerical analysis beyond the simple analytic evaluation used in the present work.
The present study simply addresses the magnetization loss on a single twisted tape, and it should be regarded as a first step toward the realistic numerical simulation for
more complicated SC high field magnets.

At last, we again emphasize that our approach in the steady state on the basis of the thin-film approximation is capable of
performing an efficient numerical evaluation of the current profile, electric field profile and magnetization loss on a SC tape surface.
A quick numerical evaluation on a SC tape surface is possible if one numerically solves equation (\ref{Faraday_current})
because it is not necessary to solve the Amp\'{e}re's law, $\bm{\nabla}\times\bm{B}/\mu_0=\bm{J}$.
In a certain limiting case, that is, in the loosely twisted tape limit ($L_{\rm p}\rightarrow \infty$),
one can readily carry out an analytical evaluation of the magnetization loss of a single twisted SC tape
without relying on a numerical approach such as finite element method.
The analytically evaluated value of the magnetization loss quantitatively agrees with the numerical value.

%\appendices

%\section{}

\section*{Acknowledgment}
%We would like to thank T. Izumi, T. Machi, M. Furuse, H. Takashima, S. Ishida (AIST), Y. Mizobata, H. Ueda, T. Tamegai, and Y. Yoshida for their discussions and valuable comments.
We would like to thank our colleagues at AIST; T. Izumi, T. Machi, M. Furuse, H. Takashima, S. Ishida, and Y. Yoshida for their discussions and valuable comments.
We also thank Y. Mizobata (Kyoto Univ.), H. Ueda (Okayama Univ.), T. Tamegai (The Univ. of Tokyo) for their valuable comments.
This work is based on the results obtained from a project commissioned by the New Energy and Industrial Technology Development Organization (NEDO).

\ifCLASSOPTIONcaptionsoff
  \newpage
\fi



%\bibliographystyle{IEEEtran}

\bibliographystyle{IEEEtran}
%\bibliography{IEEEabrv,IEEEexample}
\bibliography{IEEEabrv,180828_twisted-tape_higashi_ver2}


%\end{IEEEbiography}


%\newpage



%\vfill


\end{document}